\documentclass[preprint2]{aastex}
\usepackage{graphicx}

\def\HST{{\it HST}\/}

\def\kms{km~s$^{-1}$}

\def\gtsim{
\mathrel{\raise.3ex\hbox{$>$}\mkern-14mu\lower0.6ex\hbox{$\sim$}}
}
\def\ltsim{
\mathrel{\raise.3ex\hbox{$<$}\mkern-14mu\lower0.6ex\hbox{$\sim$}}
}
\def\farcs{\hbox{$.\!\!^{\prime\prime}$}}

\slugcomment{Submitted to the Astronomical Journal}

\shorttitle{HST Images of Cas A}
\shortauthors{Fesen et al.}
\begin{document}

\title{Hubble Space Telescope WFPC-2 Imaging of 
Cassiopeia A\altaffilmark{1}}

\altaffiltext{1}{Based on observations with the NASA/ESA Hubble Space Telescope,
obtained at the Space Telescope Science Institute,
which is operated by the Association of Universities for Research in 
Astronomy, Inc.\  under NASA contract No.\ NAS5-26555.}

\author{Robert A. Fesen\altaffilmark{2}, Jon A. Morse\altaffilmark{3}, 
Roger A. Chevalier\altaffilmark{4},
        Kazimierz J. Borkowski\altaffilmark{5},  \\ 
        Christopher L. Gerardy\altaffilmark{2}, 
Stephen S. Lawrence\altaffilmark{6}, \& Sidney van den Bergh\altaffilmark{7}  }
\altaffiltext{2}{Department of Physics \& Astronomy, Dartmouth College, Hanover, NH 03755} 
\altaffiltext{3}{Center for Astrophysics and Space Astronomy, University of Colorado, 389 UCB, Boulder, CO 80309 }
\altaffiltext{4}{Department of Astronomy, University of Virginia, P.O. Box 3818, Charlottesville, VA 22903  }
\altaffiltext{5}{Department of Physics, North Carolina State University, Raleigh, NC 27695 }
\altaffiltext{6}{Department of Physics and Astronomy, Hofstra University, Hempstead, NY 11549}
\altaffiltext{7}{Dominion Astrophysical Observatory, Herzberg Institute of Astrophysics, NRC of Canada, 
                 5071 West Saanich Road, Victoria, BC V9E 2E7, Canada  }

\begin{abstract}

The young galactic supernova remnant Cassiopeia A was imaged with WFPC-2 
aboard the {\it Hubble Space Telescope} 
through filters selected to capture the complete velocity range of the 
remnant's main shell in several emission lines. 
The primary lines detected along with the specific WFPC-2 filters used were: 
[O~III] $\lambda\lambda$4959,5007 (F450W), 
[N~II] $\lambda$6583 (F658N),
[S~II] $\lambda\lambda$6716,6731 + 
[O~II] $\lambda\lambda$7319,7330 + [O~I] $\lambda\lambda$6300,6364 (F675W), and 
[S~III] $\lambda\lambda$9069,9532 (F850LP).  About three-quarters of the 
remnant's $4'$ diameter main shell was imaged with all four filters 
in three WFPC-2 pointings, with most remaining shell regions imaged in just
the F675W filter via three additional pointings.
Considerable detail is observed in the reverse-shocked ejecta with typical knot 
scale lengths of $0\farcs2 - 0\farcs4$ ($1 - 2 \times 10^{16}$ cm).
Both bright and faint emission features appear highly clumped with little in
the way of a smooth, diffuse emission component detected.
Strong differences in [S~III] and [O~III] line intensities 
indicating chemical abundance differences are also seen,
particularly in knots located along the bright northern limb 
and near the base of the northeast jet.
A line of curved overlapping filaments in the remnant's northwestern rim
appears to mark the location of the remnant's reverse shock front in this region.
The morphology of some finger-like ejecta structures elsewhere suggest cases  
where the reverse shock front is encountering the remnant's clumped ejecta. 
Large velocity shears ($\simeq$1000 km s$^{-1}$) possibly associated with 
the formation of these Rayleigh-Taylor-like features are found in the 
line profiles of several emission lines
(e.g., [S~III] $\lambda\lambda$9069,9532 and [Cl~II] $\lambda$8679)
in ground-based, optical spectra of knots near the remnant's center.
[N~II] images of the remnant's circumstellar knots (``QSFs'') reveal them to be 
$0\farcs1 - 0\farcs6$ thick knots and filaments 
often with diffuse edges facing away from the center of expansion. 
Three color composite images of the whole remnant and certain sections
along with individual filter enlargements of selected regions of the 
bright optical shell are presented and discussed.

\end{abstract}
\keywords{ISM: individual (Cassiopeia A) - supernova remnants - 
ISM: abundances and dynamics }

\section{Introduction}

Cassiopeia A (Cas A; SN $\simeq$1680) is currently the youngest known 
Galactic supernova remnant (SNR). It is also the prototype for the class of 
young, oxygen-rich remnants containing SN ejecta moving at several thousand 
km s$^{-1}$ and exhibiting extreme O and Si-group (Si, S, Ar, and Ca) 
abundances due to explosive nuclear processing in a massive 
star \citep{vdb88}. Mass estimates for the Cas~A progenitor
range between 10 and 30 M$_{\sun}$ \citep{Fabian80,Jansen88,Vink98}.

Elevated abundances of O-burning products in the remnant's high-velocity ejecta
(``fast-moving knots'';``FMKs''), the presence of much slower moving 
He- and N-rich clumps (``quasi-stationary flocculi''; ``QSFs'') of
circumstellar material (\citealt{ck78,ck79}), together with the
presence of N-rich, high-velocity outer ejecta \citep{Fes87,Fes01} have 
prompted several researchers to suggest the Cas~A progenitor may have 
been a WN-type Wolf-Rayet star that experienced substantial mass-loss 
before exploding as a Type Ib/c or Type II 
supernova \citep{Langer86,Schaeffer87,Fes91,GS96,Vink96}.

Cas A's optical nebulosity is largely confined to a spherical 
shell $\simeq 2'$ in radius and expanding at $-4500$ to $+6000$ km s$^{-1}$.
High-velocity, radiative ejecta filaments and FMKs --- which exhibit
strong O, S, and Ar lines but no H, He, or N emission --- dominate
the SNR's optical structure and are mainly organized in  
large ring-like structures (diameters $\sim 0.5 - 1.0$ pc) situated on the 
surface of the expanding shell (\citealt{Reed91,Law95}).
Fainter optical ejecta have been found outside of the shell, 
mainly located in a northeast (NE) flare or ``jet'' of S-rich ejecta,
with some 70+ other fainter knots elsewhere around the remnant's 
outer periphery (see \citealt{Fes01} and references therein). 

The remnant's main shell of optical emission is generally believed to be 
reverse-shock heated knots of relatively dense ejecta lying some distance 
behind the outer blast wave \citep{McKee74}. Strong and complex 
X-ray, infrared, and radio emissions are also associated with the reverse shock
(e.g. \citealt{Anderson91,Lagage96,Hughes00}).  The reverse shock moving
back into the ejecta arises due to the remnant's interaction 
with a dense circumstellar medium (CSM) shell, and recent models can account 
for the observed $\sim$ 2000 km s$^{-1}$ expansion rate seen in the radio  
\citep{cl89,Borkowski96}. Evidence for deceleration of the main shell's 
optical knots can be inferred from differences between derived explosion 
dates for them (A.D. 1658 $\pm$3; \citealt{Kvdb76a}) and outer knots ahead 
of the primary blast wave (A.D. 1671 $\pm$1; \citealt{Thor2001}). 
However, high-resolution observational data tracing individual ejecta knot 
deceleration dynamics, excitation, and evolution in Cas~A are still lacking.

Recent observations have begun to shed light on related dynamical
processes in other young remnants. For example, \citet{Hester3} have
used {\it Hubble Space Telescope} (\HST) imaging observations and numerical 
models to investigate the development and ionization of magnetic Rayleigh-Taylor
instabilities at the interface between the pulsar-driven synchrotron nebula
and a shell of swept-up ejecta in the Crab Nebula. \citet{Blair00} have
successfully modeled \HST\ spectroscopic observations of UV/optical reverse
shock emission from O-rich ejecta in the young remnants N132D and
1E0102.2--7219 in the Magellanic Clouds.

To date, the fine-scale structure of Cas A's SN ejecta has been only 
moderately constrained by ground-based observations. The best published 
optical images show little easily discernible sub-arcsecond 
structure \citep{vdbP86} along the remnant's northeast 
rim while radio observations at 2~cm have resolved small $\sim0\farcs3$
features in one region \citep{AD87}. A portion of the northeast rim of
Cas~A's bright, main shell was imaged with the \HST\
Wide Field/Planetary Camera-1 (WFPC-1) prior to this study.
Two 2100~s exposures were taken using narrow [O~III] $\lambda$5007 and 
[S~II] $\lambda\lambda$6716,6731 filters. In spite of the aberrated point 
spread function of WFPC-1 and the narrow passbands of the 
filters, some filament details were detected in the low-velocity portions 
of several NE knots (V$_{\rm r}$ $\leq$ 2000 \kms)
not visible in the ground-based images \citep{vdbK85}.

Here we present the first high spatial resolution optical survey of the Cas~A 
remnant with \HST\ using filters that cover the remnant's full $-4500$ 
to $+6000$ km s$^{-1}$ expansion velocity in both intermediate- and 
low-ionization lines of sulfur and oxygen.  These images reveal a wealth of 
fine-scale spatial features in the metal-rich ejecta of this $\simeq$ 320 yr 
old SNR. During our discussion, we assume a distance to Cas~A of 
3.4 kpc \citep{Reed95}, where $1'' \approx 5 \times 10^{16}$ cm.

\section{Observations}

\subsection{{\it HST} Imaging}

We obtained multi-band images of Cas A using the Wide Field Planetary Camera-2 
(WFPC-2) aboard \HST\ during nine orbits in January 2000. 
The optically bright NE, NW, and SW rims of the main shell were targeted, 
with each imaged in four filters.  
Table 1 lists the four filters used along with their
full-width-at-zero-intensity (FWZI) bandpasses, exposure times used, and
principal line emissions within the bandpass.
Exposure times ranged from 400~s to 700~s and images were taken in pairs
of two slightly offset exposures in order to facilitate removal
of cosmic ray hits and dead pixels in the individual CCD images.

Because of the large 
expansion velocity of the remnant's bright shell of ejecta,
we employed broad rather than narrow passband WFPC-2 filters, including 
F450W, F675W (the \HST\ R band), and F850LP. 
In addition to these, images were taken using the narrow passband F658N filter 
which is sensitive to the remnant's slow-moving, 
[N~II] $\lambda\lambda$6548,6583 emitting circumstellar material. 
The resulting images provide a high-resolution, four-passband survey of the 
remnant's major optical features for these three remnant sections. 
Adjacent remnant shell regions to each of the three main target positions 
were also obtained in just the F675W filter in order to give a more complete 
survey of Cas~A at high resolution within the granted spacecraft time.

We selected the F450W and F850LP filters to provide a comparative [O~III] 
{\it versus} [S~III] emission survey of Cas A's main shell.
The F450W and F850LP filters effectively isolated ejecta line emissions of 
[O~III] and [S~III], respectively, while covering the full $\pm 6000$ \kms  \
velocity range. The WFPC-2 Instrument Handbook shows that the throughput of the 
F450W filter + telescope + detector system peaks near the 
[O~III] $\lambda\lambda4959,5007$ lines and
covers an [O~III] radial velocity range of more than $\pm 8000$ km s$^{-1}$. 
The filter is only weakly sensitive to [O II] $\lambda\lambda$3726,3729 
line emission at about 10\% the transmission at 5007 \AA. Significant 
contamination from the [S~II] $\lambda\lambda$4068,4062 lines is 
unlikely since these lines are weak in most FMKs \citep{KC77,ck79,Hur96}.

The long-pass filter F850LP (8300 -- 10600 \AA) 
was similarly chosen to isolate the remnant's [S~III] emissions. 
Its bandpass, together with WFPC-2 CCD sensitivity steep decline at 
wavelengths $>$ 9500 \AA, makes it is primarily sensitive to ejecta 
[S~III] $\lambda\lambda$9069,9531 line emission, and only about 10\% 
as sensitive to the [S~II] $\lambda\lambda$10290-10370 line blend.
Ground-based spectra show that contributions from [Fe~II]$\lambda8617$
and [C~I]$\lambda\lambda9824,9850$ are also likely to be small.

We chose to use the broad ``R band'' F675W filter (6000 -- 7500 \AA)
for imaging lower ionization line emission features which exhibit strong lines
of [O~II] $\lambda\lambda$7319,7330 and [S~II] $\lambda\lambda$6716,6731.
Although this filter's bandpass also covers the [O~I] $\lambda\lambda$6300,6364
and [Ar~III] $\lambda$7135 lines, both are much weaker than either the 
[O~II] and [S~II] lines for most filaments \citep{ck79,Hur96}. 
To help discriminate O,S,Ar-rich ejecta from the 
[N~II] $\lambda\lambda$6548,6583 strong, circumstellar material (QSFs) detected
in the F675W images, we also obtained separate exposures through the F658N 
filter.

Details of the WFPC-2 data reduction procedures we followed
are described in \citet{Morse96} and \citet{Blair00}.
Using IRAF/STSDAS\footnote{IRAF is distributed by the
National Optical Astronomy Observatories, which is operated by the Association
of Universities for Research in Astronomy, Inc.\ (AURA) under cooperative
agreement with the National Science Foundation. The Space Telescope Science
Data Analysis System (STSDAS) is distributed by the Space Telescope Science
Institute.} software tasks, the pipeline-calibrated exposure pairs through each
filter were carefully aligned and combined to reject cosmic rays.
Stray hot and dead pixels were then flagged and corrected. 
We applied geometric distortion corrections to each CCD chip using
the Trauger wavelength-dependent coefficients contained in the STSDAS.DITHER 
package and then flux calibrated using the methods described in the
WFPC-2 Instrument Handbook. 

Finally, we applied an extinction correction to each image, appropriate
for the rest wavelength of each emission line observed, that was based on
an average color excess of E(B$-$V) $\approx 1.65$ \citep{Hur96}
and the extinction curve of \citet{Card89} with
R$_{\rm V}$ = 3.1. The extinction across the face of Cas~A is known to vary,
as seen both in optical/NIR spectra (e.g. \citealt{Hur96}) and a
neutral H column density measurement \citep{Keohane96}.
The highest neutral H column density observed is on the western limb.
This region contains bright X-ray and radio emission, but very little optical emission.
As a result, and because we are not making detailed abundance estimates
based on the emission-line image ratios, we applied a representative
extinction correction for the optically emitting filaments. However,
large-scale gradients in the extinction may account for the somewhat
redder appearance of the emission along the SW rim in the color composite
of Figure 1.

With the cleaned, fully calibrated images in hand, 
a large contiguous mosaic was built from the six individual
pointings through the F675W filter. These six pointings generally overlapped
sufficiently to provide several stars that could be used as tie points
in co-aligning adjacent exposures. Because we found that using the
STSDAS task WMOSAIC yielded measurable errors in the inter-chip rotations
and offsets when piecing the large mosaic together, we instead rotated
each Wide Field (WF) camera chip so that North was toward the top
using the astrometry information in the individual chip science headers
and independently determined the inter-chip rotations and offsets. 
In regions of the SNR where adjacent images overlapped, the data were averaged.

After assembling the large F675W mosaic of WF chips, we then added in the
Planetary Camera (PC) chip images in regions where they provided additional
coverage of the SNR emissions. We de-magnified the PC pixel scale by
a factor of 0.4555 to match the $0\farcs1$ WF pixel scale and inserted the 
re-scaled PC images into their appropriate positions. Finally, we used the
F675W mosaic as the common reference frame for aligning the individual chips 
of each F450W, F658N, and F850LP image at the three primary pointings.
The resulting mosaic images, as well as some full-resolution PC frames,
are discussed in Section 3.

\subsection{Spectroscopy}

As part of exploratory ground-based observations prior to these \HST\ 
observations, low-dispersion optical spectra of several ejecta knots north 
of the remnant's center were obtained in October 1998 using the 
Hiltner 2.4~m telescope at the MDM Observatory. 
A Modular Spectrograph was used together with a 600 lines mm$^{-1}$ grating 
blazed at 5000 \AA\ and an east-west $1\farcs5$ wide slit. 
Several 1200~s exposures were obtained with a wavelength coverage of 
6300 to 9800 \AA\ and a spectral resolution of 7 \AA. These data were 
reduced using standard IRAF software routines and calibrated with 
Hg, Ne, and Xe lamps and \citet{Massey90} standard stars.

\section{Results and Discussion}

A three color composite WFPC-2 image covering most of the Cas A SNR is shown 
in Figure 1.  The remnant's appearance in the F850LP, F675W, and F450W 
images is shown in red, green, and blue, respectively. Areas which were 
imaged only in the F675W filter are shown in grayscale. With the exception 
of the westernmost portion where little optical emission is present, the
WFPC-2 imaging covered Cas A's entire bright optical shell.

The presence of line emission variability across the remnant is 
visible as color differences in Figure 1. Both ionization structure and
chemical fractionation manifest themselves as color variations.
Examples include a few blue [O~III] knots along the northeast limb,
a blue-green emission arc along the southern portion of the remnant's
bright northern rim, and reddish filaments and knots in the southwest.
Even at the relatively low resolution of this reproduction, 
considerable small-scale detail can be discerned that is not visible in 
previous ground-based images (see false-color image in \citealt{Kvdb76b}).

In the following subsections, we briefly describe the remnant's 
basic emission properties and structures as seen in these \HST\ images. 
A more thorough investigation of the remnant's optical 
properties and varying knot chemical abundances making use of spectroscopy and 
detailed shock modeling will be addressed in a subsequent paper.

\subsection{Main Shell Structure}

In the best published ground-based optical images, 
Cas A's optical shell appeared composed largely of $1'' - 10''$ diffuse 
clumps and filaments (\citealt{vdbK85,vdbP86}). 
However, the WFPC-2 images reveal a far more complex ejecta structure,
one rich in fine-scale features.  We begin our discussion by 
examining two of the remnant's brightest limb regions.

The complexity of the remnant's optically emitting ejecta can be seen in two 
of the remnant's brightest regions: the large Baade-Minkowski Filament 
No.\ 1 (\citealt{BM54}) situated along the NE limb, and a large arc of 
nebulosity bordering the southwestern rim.  Figure 2 shows 
enlargements of F675W filter images of both these regions. This filter is 
principally sensitive to relatively low ionization line emissions including 
[S~II] $\lambda\lambda$6176,6731 and [O~II] $\lambda\lambda$7319,7330, 
and the generally weaker [O~I] $\lambda\lambda$6300,6364.
It is also sensitive to [Ar~III] $\lambda$7135 emission but this
is typically a faint line \citep{KC77,ck79,Winkler91,Reed95,Hur96}. 

The Baade-Minkowski Filament No.\ 1 image (upper panel) was taken using 
the PC and therefore has about twice the resolution of the WF image 
taken of the SW rim  (lower panel).  Nonetheless, a wealth of detail is 
seen in both regions. Filament No.\ 1 is resolved
into an array of bright knots and short filaments $\sim 0\farcs2 - 0\farcs6$ in size. 
Much of the fainter emission detected appears also to be highly clumped, 
with little in the way of smooth, diffuse emission present.

The same general morphology is seen in the remnant's southwestern
rim nebulosity, but with more radially oriented emission features. 
Again numerous small knots ($\leq 0\farcs5$) are present with a clumpy, faint
inter-knot background.  The brightest feature is a vertical cloud structure
near the western edge of the region shown.  This is not SN ejecta but a clump
of circumstellar mass-loss material (QSF R9; see Section 3.6 below).

It is instructive to compare these images with ground-based images presented 
in \citealt{vdbK85} (their Figs. 9 \& 11) which also show their size and 
brightness evolution over the period 1958 to 1983. What appear to be diffuse
clumps and short filaments several arcseconds in size on these photographs 
are resolved now into clusters of numerous sub-arcsecond features.

Figure 3 shows false-color images of these regions with the same color-coding 
used in Figure 1. In making these figures, we enhanced the colors over those of 
the observed line intensity ratios to increase the visibility of emission 
differences.  Features that are roughly equally bright in all three bands 
appear white to slightly yellow in color and correspond to a ``typical'' 
Cas~A ejecta spectrum; i.e., having strong 
[S~III] $\lambda\lambda$9069,9531,
[S~II] $\lambda\lambda$6716,6731, [O~III] 
$\lambda\lambda$5007,4959, and [O~II] $\lambda\lambda$7319,7330.
One such example is the bright white-ish feature along the top western portion 
of the Baade-Minkowski Filament No.\ 1 for which spectra were obtained 
by \citet{Hur96}.

Although some sections of both regions appear dominated by one of the three
filter images (blue for [O~III] or red for [S~III]), there does not appear to 
be a systematic pattern or arrangement to these regions. Small clumps of blue 
[O~III] bright knots, greenish QSFs (i.e., detected only in the F675W filter) 
and red [S~III] clumps appear scattered throughout each region
with no obvious ionization structure either on large or small scales.

Organized emission variations are, however, present in some other regions
of the remnant.  This is shown in Figure 4 which presents a section of the
remnant's north-central shell. Color differences between features have been
again enhanced.  The two blue-green arcs along the southern portion are in
sharp contrast to several reddish appearing regions lying to the north and
the blue-green knots near the image top.  The bluest and reddest features
in this region reflect significant [O~III] to [S~III] line intensity 
differences, suggestive of O/S abundance variations like those found 
by \citet{ck79}. 

For example, a spectrum for the yellowish, peak-shaped feature at the 
northern junction point of the two southern arcs shows unusually strong 
[O~II] $\lambda\lambda$7319,7330 emission 
(``FMK4''; [O~III]:[O~II]:[S~III]:[S~II] = 4:14:11:1;  \citealt{Hur96}).
It important to bear in mind that these are observed relative line
strengths, and [O~III] emission (not [S~III]) is most often the strongest
optical line emission once extinction corrections are applied \citep{Hur96}. 

In a similar fashion, the pinkish colored regions lying just 
above the bright star near the left-central portion
of the image exhibit spectra in 
which [S~III] $\lambda\lambda$9069,9531 dominate the optical/NIR
line emissions (``FMK2'' \& ``FMK3''; 
[O~III]:[O~II]:[S~III]:[S~II] = 4:4:18:3; \citealt{Hur96}). 
Even stronger [S~III] emission regions are found
above the double star in the lower right (``FMK5''; 
[O~III]:[O~II]:[S~III]:[S~II] = 4:6:32:6; \citealt{Hur96}) 
and consequently appear here significantly redder.

Interestingly, the large oval shaped emission structure only partially seen
here (but see Fig. 1), bounded on the south by the blue-green [O~III] bright arcs,
and on the north, east, and southeast by red [S~III] bright emission
appears kinematically to be a single, coherent ring-like structure.
When sampled via its [S~II] $\lambda\lambda$6716,6731 emission, this ring
spans a radial velocity range from $-2700$ km s$^{-1}$ in the southwest to $-900$ km s$^{-1}$
in the north with a characteristic mean around -1800  km s$^{-1}$ \citep{Law95}.
The ring's most blueshifted material ($-2700 \pm 300$ km s$^{-1}$) 
is the orangish-colored features near the ``peak-like'' filament 
at the bottom of the ring and extends westward to just south of the bright double stars. 
The radial velocity distribution of the [S~II] emitting gas increases smoothly as one
moves around the circumference of the ring-like structure,
ending at around $-900$ km s$^{-1}$ in the fainter knots in the
northern half, directly above (north) of the ``peak'' feature.  

\subsection{Ejecta Knot Dimensions}

Most of Cas~A's optical ejecta shell consists
of clumpy nebulosities with angular knot dimensions of $\leq$ $1''$ 
The presence of clumpy ejecta has already been briefly
discussed above in regard to the large emission regions shown in Figure 2.
There, as elsewhere throughout the remnant's optical emission features,
a majority of ejecta knots have a scale-length for the smallest features
of some $0\farcs2 - 0\farcs4$ ($1 - 2 \times 10^{16}$ cm) in diameter.
Interestingly, this size is nearly the same as that
determined from high-resolution 2~cm radio
observations of the remnant which found 
$\simeq 0\farcs3$ knots for one small region \citep{AD87}.

A clear illustration of the clumpy nature of the remnant's ejecta
can be seen in Figure 5 which shows a small portion of the remnant's northern
limb of bright and extensive optical emission. Here we show
the WFC F675W image of the region magnified
by a factor of 2 in order to lessen the appearance of individual pixels.
Dozens of bright $0\farcs2 - 0\farcs4$ size knots arranged
in $\sim10''$ (0.15 pc; $5 \times 10^{17}$ cm) long chains can be seen in this region.
Much of the remnant's northern ``wreath'' of optical
emission, in fact, breaks up into clouds of
tiny knots with angular dimensions mostly $\leq 0\farcs5$.
Although no stars were removed from any of the images
(since no pure continuum, off-band images were taken),
there are very few faint stars visible in the field-of-view shown.
So nearly all of the small unresolved sources seen in this figure are
SN ejecta knots.

Here, as was the case for the bright emission regions examined in Figure 2,
fainter emission features in between and around the brighter knots
appear about as clumpy and non-uniformly distributed as the brighter features.
In addition, much of the fainter emission in this 
area is associated with the brighter
knots, often located along the inner edges of the knotty loops and strings.
Emission features, both bright and faint,
have structure right down to the image resolution ($0.1''$; $5 \times 10^{15}$ cm).
Examination of higher resolution PC images suggest that while there is considerable
fine structure in the knots and short filaments undersampled by the WFC images,
the basic knot scale length throughout the optical shell is consistently on the
$0\farcs2 - 0\farcs4$ size scale.

Supernova models for intermediate- and high-mass progenitors 
suggest that ejecta will form clumps
at the interfaces of layers of different chemical composition
chiefly due to Rayleigh-Talyor (R-T) instabilities 
\citep{Chevalier76,Arnett89,Fryxell91,Yamada90,
Herant91,Herant92,Hachisu94}. Such R-T ejecta knots may then undergo strong
compression after their passage through the reverse shock
and into the reverse shock - front shock region \citep{Cid-Fernandes96}.
However, R-T instabilities depend upon internal presupernova structure
and little is currently known empirically concerning specific ejecta knot 
size and evolution following the reverse shock.

\citet{Hachisu91} and \citet{Herant94} have shown that their models of the 
post-explosion hydrodynamics
in intermediate mass progenitors (12--30 M$_{\odot}$) exhibit a highly clumpy ejecta
structure in the post-reverse shock region where R-T instabilities become chaotic.
The model simulations of \citet{Herant94} give results that resemble the Cas~A jet knots
\citep{Fes96}. However, due in part to grid-mesh limitations,
it is not clear if the knot scaling in such models is close to that
actually observed either in the jet or in the main shell.
Nearly all models show R-T fingers and associated features at least an order
of magnitude larger than the typical knots scale-lengths seen in Cas~A.
The degree of any ejecta clumping in post-reverse shock region is
also likely to be increased by thermal instabilities from enhanced collisional
cooling \citep{Herant94,SD95}.
The O-rich shock models of \citet{Blair00} show post-shock compression
factors $>100$ in the fully radiative knots.
  
\subsection{Chemically Peculiar Knots}

Previous spectra and imaging found several O-rich filaments  
situated along the eastern limb near the base of the remnant's 
NE ``jet'' \citep{Kvdb76b,ck79}.
These characteristically exhibit strong [O~III] emissions
and thus show up as strongly blue features in our three color
images of this region (Fig.\ 6). The brightest, bluest features visible in 
Figure 6 turn out not to have been present
three decades ago, while others visible in the early 1970's 
(e.g. the most eastern blue filaments seen in Fig.\ 6) 
have significantly faded. 

Several seemingly S-rich but O-poor ejecta knots have also been identified. 
One such knot was the NW knot KB~33 \citep{Kvdb76a} studied by \citet{ck79}.
This knot has unfortunately faded and today is not readily identifiable.
A somewhat more extreme S-rich knot was discovered by \citet{Hur96}
but was imaged with the WFPC-2 only using the F675W filter.
However, several [S~II] strong knots have been found
in and near the base of the NE jet \citep{Fes96},
and some of these are seen as deep red knots in Figure 6.

Since color images do not always accurately illustrate the
strength or weakness of some features on the individual images,
we show in Figure 7 the F450W and F850LP for this same eastern limb
region. Here the blue features visible in Figure 6 appear to have
virtually no [S~III] emission, whereas the [S~III] (F850LP) strong red knots  
show little or no [O~III] (F450W) emission. Both O-rich and S-rich types of 
features do show up, of course, in the F675W image presumably due to
either the presence of other oxygen or sulfur lines respectively.

Because [O~III] and [S~III] have relatively high ionization energies
(54.9 eV and 34.8 eV, respectively) and similar electronic structures,
it seems unlikely that such gross line ratio differences can be accounted
for by means other than abundance differences.
Therefore, the presence of morphological differences between the types of knots,
where O-bright knots appear more nebulous than the more knotty S-rich ones (see Fig. 6),
may reflect the physical processes which create ejecta knots in various 
progenitor layers. 

\subsection{Large-Scale Reverse Shock Features}

The remnant's bright radio and X-ray shell structure has been generally 
taken as marking the location of the reverse shock front \citep{Gotthelf01}.
Until now there has been little evidence in the remnant's optical shell
structure to indicate the presence of an organized reverse shock front.
However, the {\it HST} images reveal some optical emission 
structures that appear to be associated with this shock.

A portion of the northwestern corner of the remnant shell is shown in Figure 8.
Despite the center of expansion lying some distance off the lower left hand 
side of the top panel image, a line of overlapping, concave curved emission 
filaments can be seen (they appear reddish in color in Fig.\  1). 
The structure is similar to that commonly seen in 
older SNRs except here the expansion orientation is reversed in the sense of 
the implied motion of the filament
curvature is towards rather than away from the remnant center.

On ground-based images, this line of emission appeared as nothing more than a 
series of medium size ejecta knots and filaments with little evidence for any 
morphologically coherent structure (Fig. 10; \citealt{vdbK85}). 
However, velocity maps in the [S~II] doublet show that this line of 
emission is dynamically quite coherent,
with velocities systematically decreasing +900 to +600 km s$^{-1}$ 
from north to south \citep{Law95}. 
The cluster of yellowish knots located farther west along the southern fringe of 
this line of filaments (see Fig.\ 1) show considerably larger redshifts 
(+1200 to +1700 km s$^{-1}$) and thus are separate, unrelated features.

The morphology of the overlapping filamentary structure shown in Figure 8 
suggests it marks the location of the 
reverse shock front in this region of the remnant. Here, as elsewhere, 
the fine-scale emission structure again is highly clumped 
especially radially away from the leading (inner) edge
of emission. Ejecta knot dimensions range 
from $0\farcs2 - 0\farcs6$ with a
strong radial asymmetry away from the shock front.  
A higher resolution PC close-up image of a portion of this 
region is shown in the lower panel
of Figure 8. Ejecta features down to the PC's $0\farcs045$ resolution 
($2 \times 10^{15}$ cm)
can be seen both at and away from the shock front.

Orientation of these features is generally nonradial with respect to the
explosion center, and varies systematically along the shock. This strongly
suggests the presence of nonradial shear flows behind the shock, which are
expected to occur behind the oblique reverse shock seen in this
region of Cas A. A test of the reverse shock nature of
these filaments may come with a second epoch of {\it HST} images of the region. 
If the filament edge does indeed mark the
location of the reverse shock front, its proper motion outward from
the remnant center should be somewhat less than that of individual ejecta knots.

In a wider view of this region as seen in the F675W filter,
a faint, long ($\sim15''$; 0.25 pc) thin filament lies just off 
to the east of these bright curved filaments (Fig.\ 9).
This filament has a thickness comparable to that of the WFC resolution ($\leq 0\farcs2$)
with some exceedingly faint diffuse emission extending some $2'' - 5''$ 
($1 - 3 \times 10^{17}$ cm) to the east. The filament shows up weakly in the [S~III] F850LP 
filter image but not in the F450W image.
Though faint (e.g. it is not visible in Fig.\ 8),
it is however definitely real and is seen in deep, 
ground-based R-band images of the SNR.

The filament's smooth, continuous appearance is singular 
within the remnant, with nothing else like it morphologically 
in the F675W or the other three filters. 
Due to its projected location adjacent to 
and closer in to the Cas~A expansion center
relative to the line of the suspected reverse shock emission discussed above, 
it is tempting to associate
it somehow with the brighter structure, possibly as shock 
photoionization precursor emission or a faint
reflection of neighboring FMK emission by ejecta dust \citep{Douvion01}.
Due to its extreme faintness, however, high quality spectroscopic observations 
needed to determine its true nature may prove difficult to obtain.

Other remnant features suggestive of reverse shock front and associated 
instabilities are shown in Figure 10. 
In the upper left hand panel, we show an
emission feature located in the south-central portion of the remnant.
A R-T-like ``head-tail'' feature with a bright knot
at the tip of a thin stem appears strikingly like that of a model calculations
of a shock running over a dense cloudlet (\citealt{Klein94}, \citealt{Klein00}).
Its projected location in front of (to the southeast; lower left) 
of a faint scalloped line of emission
is quite suggestive of shock front emission (possibly that of the reverse shock)
moving back towards the remnant center (northwest; upper right).

Finer and more closely packed R-T-like features are present
along the remnant's eastern limb, near the jet breakout region (Fig.\ 10, 
right panel).  Here one sees a series of small $\sim 0\farcs2$ knots at the ends 
of finger-like filaments which in turn appear connected to a line of much 
brighter emission knots. 
These filaments are markedly nonradial with respect to the explosion center,
again hinting at the presence of shear flows behind the reverse shock.
Lines of closely spaced, bright knots that have 
trailing emission are a common feature of the shell (Fig.\ 8).

\subsection{Ejecta Knot Mass Stripping}

The fine-scale structures of the remnant's ejecta suggest partially fragmented
knots with some mass stripping possibly due to Kelvin-Helmholtz instabilities along
knots edges like that modeled by \citet{Klein94}. Relatively large velocity dispersions
seen in individual knots \citep{vdb71,ck79} have been attributed
to the reverse shock deceleration but there has been no clear evidence for this optically.
However, we have detected spectroscopic evidence for a large velocity shear
that may help in our understanding of some of the knot features seen 
in the {\it HST} images.

In exploratory long slit spectra taken of emission regions located 
just north of the remnant's expansion center,
a few knots were found to exhibit asymmetrical emission ``streaks'' $\simeq$ 1000 km s$^{-1}$
from line centers. One especially bright knot showing this effect 
is located in the northwest portion of the remnant.
The slit position across the remnant that detected this knot 
is shown in Figure 11. Several emission knots (Nos.\ $1 - 5$)
were detected with the brightest one being Knot~5 located near the slit's western end.
The bright knot's line emission is actually a blend of two neighboring, large
emission clumps (Fig.\  11, upper right hand panel)  with the more eastern one being
considerably brighter than the more western one.
All five emission knots exhibit large redshifted emission ($>$ 4000 km s$^{-1}$)
placing them on the remnant's receding hemisphere.  

Several emission lines are shown in the lower four panels of Figure 11
including [S~III] $\lambda\lambda$9069,9531 , [S~II]$\lambda\lambda$6716,6731, 
[Cl~II] $\lambda$8579 and [Fe~II] $\lambda$8617.  
The [S~III] lines show a conspicuous blue-shifted emission ``streak''
extending roughly 1000 km s$^{-1}$. 
The peak intensity level of the streaked emission is about 10\% of
the peak [S III] line emission.
No similar blue (or red) shifted emission feature is seen in the [S~II] line
down to a level $\sim $1\% of the line's peak emission.
However, a blue shifted emission feature similar to that seen in the [S~III]
lines is present for other high ionization lines including [Cl II]
and [Ar III] $\lambda$7135. If interpreted as true knot deceleration, this 
$\sim$1000 km s$^{-1}$  velocity spread represents a $\sim$ 20\% velocity
decrease; that is, from an observed radial velocity 
of +5150 to around +3800 km s$^{-1}$. Such velocity shearing may be
related to some R-T-like features seen in the remnant.

\subsection{Circumstellar Knots (QSFs)}

In their discovery of optical emission associated with Cas A, 
\citet{BM54} noted the presence of several ``red bits'' of nebulosity 
which showed imperceptible proper motions and small, 
mostly negative radial velocities.
Subsequent work showed these features, for which \citet{Minkowski57} coined 
the term ``quasi-stationary flocculi'' or ``QSFs'', 
to have optical spectra dominated by 
[N~II] $\lambda\lambda$6548,6583 and H$\alpha$ line emission.
Due to their low expansion velocities ($\mid v \mid$ $\leq$ 500 km s$^{-1}$) and 
[N~II] 6583/H$\alpha$ ratios $\simeq$ 3 indicative of N/H enrichment, 
QSFs are believed to be clumps of circumstellar mass loss material shocked
by the remnant's blast wave \citep{vdb71,Pvdb71,Kvdb76a,vdbK85}.     

About 40 bright QSFs have been identified to date. 
They are non-uniformly distributed around the remnant, with
concentrations along the north and southwest limbs,
plus a handful found outside of the main shell in the SW
(see \citealt{vdb71}, \citealt{vdbK85}, \citealt{Law95}, \citealt{Reynoso97}, and \citealt{Fes01} 
for images and ID charts; Note: \citet{Reed95} reported the detection of 454 QSFs 
but quote a [N~II] $\lambda$6583 detection flux limit 
nearly an order of magnitude above that of the deepest H$\alpha$ + [N~II] images).

QSFs are irregular or elongated in shape with
typical diameters of  $1'' - 5''$ \citep{min68,vdb71}.
Two morphological QSF classes have been proposed by
\citet{vdbK85} consisting of: 1)
elongated, head-tail or ``tadpole-like'' objects mainly found in southern regions, 
and 2) smaller, more compact clumps typically found in the northern
FMK-rich areas \citep{vdb71}. 
They also suggest that tadpole shaped QSFs 
might be older clumps of shocked CSM than the more compact QSFs.

WFPC-2 [N~II] F685N images of nine representative QSFs from different sections around
the remnant are shown in Figure 12. QSF knot identification numbers are those
of \citet{vdbK85} with the exception of ``R41'' which we use to denote 
the cluster of three semi-stellar QSFs northwest of 
R5 not identified by \citet{vdbK85}.
QSFs from the northern half of Cas A's bright optical shell are
shown in the top two rows of Figure 12,
with the bottom row showing the southern knots R4, R9, and R19.

From these {\it HST} images,
QSF emission cores seem generally to be $0\farcs1 - 0\farcs6$ in size, 
with considerable associated diffuse nebulosity in many cases.
Tadpole-like objects break up into a series of individual
knots rather than being continuous features (e.g. R4 \& R23).   
Other QSFs that appear largely filamentary on ground-based images
are now resolved into a series of small knots and diffuse clumps (e.g. R17 \& R23).

In contrast to the generally more diffuse morphology of southern QSFs (R4 \& R9),
some northern rim QSFs (R26, R27 \& R41), have small
internal structures with knot sizes down to the $\sim0.1''$ WFC resolution.
While compact QSFs may be more common in the northern FMK-rich areas, it is
unclear whether such morphological differences are physically meaningful.

However, we do find that the diffuse emission associated with northern limb QSFs
lies mainly at greater radial distances from the remnant's expansion
center (e.g. R5, R7, R10-R29, \& R27). 
This is to say, the brightest internal knot features typically lie
along the inner QSFs edges, closer to the remnant's explosion center.
This could indicate partial knot disruption and mass stripping from cloud edges 
following main shock front passage like that suggested 
by some shock models \citep{MacLow94,Klein94}.  

\subsection{Cas A's General Optical Structure}

The overall appearance of Cas A is the result of several factors
including the structure of the ejecta, the structure of the ambient
medium, and the complex interaction between them on large and small scales.
The remnant's optical ejecta, the so-called FMKs, may represent only a small
fraction of the total ejected material. 
Relatively smooth, cold ejecta will pass through a reverse shock front
and get heated to X-ray emitting temperatures.
Denser clumps that are initially co-moving with the cold ejecta
will become compressed by the high thermal pressure and
the ram pressure in the postshock region leading to the optical FMK emission seen.

\citet{Reed91} and \citet{Law95} found that the FMKs tend to appear
as circular features on the surface of a sphere.
Our images tend to reinforce that appearance but show much 
finer ejecta detail (e.g. Fig.\ 5) along with whole regions
that are strongly filamentary (e.g. Fig.\ 8).
The origin of some of these curved structures and filaments may
be due in part to a ``Ni bubble effect'' 
acting in the days after the initial explosion.
\citet{Li93} showed how this effect can account for
the high volume filling factor of Fe in SN~1987A despite
its small mass, and also the clumpiness of the O emission.
Non-radioactive material can end up being compressed by rising and expanding  
bubbles of ejecta rich in radioactive $^{56}$Ni.

\citet{Basko94} followed the initial expansion of a single Ni bubble,
and \citet{Blondin01} simulated the interaction
of multiple bubbles with an ambient medium.
The presence of non-radioactive O and S in the fast knots is consistent
with this scenario, as is the presence of more diffuse Fe-rich ejecta in
X-ray emitting gas \citep{Hughes00}.
The new {\it HST} observations show detailed structure in the fast ejecta,
some of which might be the result of the initial Ni bubble formation.

The images give the impression of planar or generally slightly concave
features which may be bubble ``walls'' with sizes up to 
half of the radius of the reverse shock.
In hydrodynamical simulations of \citet{Blondin01},
the slowest shocks, which are most likely to be seen at optical wavelengths,
are generated by an impact of relatively fast shocks propagating through
the Fe bubbles with the dense bubbles' bottoms and side walls. This results
in a generally concave morphology of such slow shocks, while the interface
regions between adjacent bubbles may produce dense, outward-facing protrusions
where conditions are even more favorable for
radiative cooling and production of optical emission. The dense and cool
shocked ejecta seen in these simulations are less decelerated than the much
hotter X-ray emitting gas comprising the bulk of the shocked ejecta, in
qualitative agreement with the much higher space velocities of optically
emitting material in reverse shocks when compared with the velocity of
the bright Cas A shell inferred from X-ray data.

The nature of the ambient medium into which Cas~A is expanding
is also uncertain. Expansion into a uniform medium is often assumed, 
with the X-ray bright region representing the reverse shocked gas and the outer plateau
representing the forward shocked gas (e.g. \citealt{Fabian80};
\citealt{Gotthelf01}).
However, the inference that the progenitor of Cas~A was a Wolf-Rayet
star strongly suggests that the region around the supernova was
modified by mass loss from the progenitor.
\citet{cl89} noted that the N-rich QSFs, which are likely
to be mass lost from the progenitor, could be the denser parts of
a presupernova circumstellar shell at the radius of the strong emission.
The positions of the 40 brightest QSFs are consistent with the strong interaction
region occupied by most of the FMKs.

\citet{Borkowski96} carried out 1D simulations
of a circumstellar interaction model and modeled the observed X-ray spectrum.
In this scenario, the FMKs were brightest where they moved through
the circumstellar shell because of the high ram pressure.
But some portions of the Cas A shell exhibit little or no FMK emission.
It is possible that these optically faint regions are where the cooling time for the
shocked knots is longer than the remnant age, so the knot material may emit
in X-rays, but not at optical wavelengths.
The {\it Chandra} X-ray image (e.g. Fig.\ 1 of \citealt{Hughes00})
shows a cellular structure that is more widespread than that seen
in the optical. However, with the exception of the western rim where there is 
bright X-ray but little optical emission,
one does find a rough correspondance with areas having bright optical
knots and strong X-ray emission \citep{Fes01}.

\section{Conclusions}

The high resolution WFPC-2 \HST\ images provide 
the first look at the fine-scale structure
of metal-rich ejecta in this young SNR from a core-collapse
SN. Our main conclusions include:

1) Considerable fine-scale detail is observed in the 
reverse-shocked ejecta with typical knot
scale lengths of $0\farcs2 - 0\farcs4$ ($1 - 2 \times 10^{16}$ cm).
Both bright and faint emission features appear highly clumped with little in
the way of a smooth, diffuse emission component detected.
The origin of some curved structures and filaments may
be due in part to bubbles of radioactive Ni-rich ejecta
compressing non-radioactive material into planar-like features. 

2) Strong differences in [S~III] and [O~III] line intensities
indicating large chemical abundance differences are seen,
particularly in knots located along the bright northern limb
and near the base of the northeast jet.

3) A line of curved, overlapping filaments in the remnant's northwestern rim
appears to mark the location of the remnant's reverse shock front in this region.
The morphology of some finger-like ejecta structures elsewhere suggest cases
where the reverse shock front is encountering the remnant's clumped ejecta.
A faint filamentary emission structure is found to lie some 
5$''$ east of the line of suspected reverse shock front filaments,
and might be shock precursor emission in more inward-lying ejecta
or a reflection of neighboring FMK emissions off ejecta dust.

4) Large velocity shears ($\simeq$1000 km s$^{-1}$) possibly associated with
the formation of Rayleigh-Taylor-like features are found in the
line profiles of several emission lines
(e.g. [S~III] $\lambda\lambda$9069,9532 and [Cl~II] $\lambda$8679)
from ground-based optical spectra of knots near the remnant's center.
Similar velocity shearing in knots may be
related to R-T-like features seen in other portions of the remnant.
The nonradial orientation of several well-defined systems of linear filaments
hints at the presence of large-scale shear flows behind the reverse shock.

5) [N~II] images of the remnant's circumstellar knots (``QSFs'') reveal them to be
$0\farcs1 - 0\farcs6$ thick knots and filaments with diffuse edges facing away from
the center of expansion.

The evolution and proper motion of the remnant's fine-scale ejecta features
along with possible ionization changes will be investigated
through follow-up WFPC-2 imaging.

\acknowledgments

We thank Zolt Levay at STScI for assistance with the preparation
of the color images, and the referee for several helpful comments which improved
the clarity of the text. Support for this work was provided by NASA through grant number GO-8281 
from the Space Telescope Science Institute,
which is operated by AURA, Inc., under NASA contract NAS5-26555.

\clearpage

\begin{deluxetable}{llllc}
\tabletypesize{\small}
\footnotesize
\tablecaption{WFPC-2 Filters and Detected Line Emissions}
\tablewidth{0pt}
\tablehead{
\colhead{Filter}  & \colhead{Bandpass} & \colhead{Primary Lines} & \colhead{Secondary Lines}
 & \colhead{Exposure Times}
}
\startdata
 F450W & 3700--5200 \AA  \ & [O III] 4959,5007                  & [O~II] 3726,3729            & $4 \times 700$ s \\
 F658N & 6560--6620 \AA  \ & [N II] 6583                        & H$\alpha$ 6563              & $4 \times 400$ s \\
 F675W & 6000--7600 \AA  \ & [S II] 6716,6731; [O II] 7319,7330;& [N II] 6583; H$\alpha$ 6563;& $4 \times 500$ s \\ 
       &                   & [O~I] 6300,6364                    & [Ar III] 7135               &                  \\
 F850LP& 8300--10600 \AA \ & [S III] 9069,9531                  & [S~II] 10287-10370          & $4 \times 700$ s \\
\enddata
\end{deluxetable}

\clearpage
\newpage

%
%

\clearpage

\begin{figure*}
\begin{center}
\includegraphics[scale=0.85]{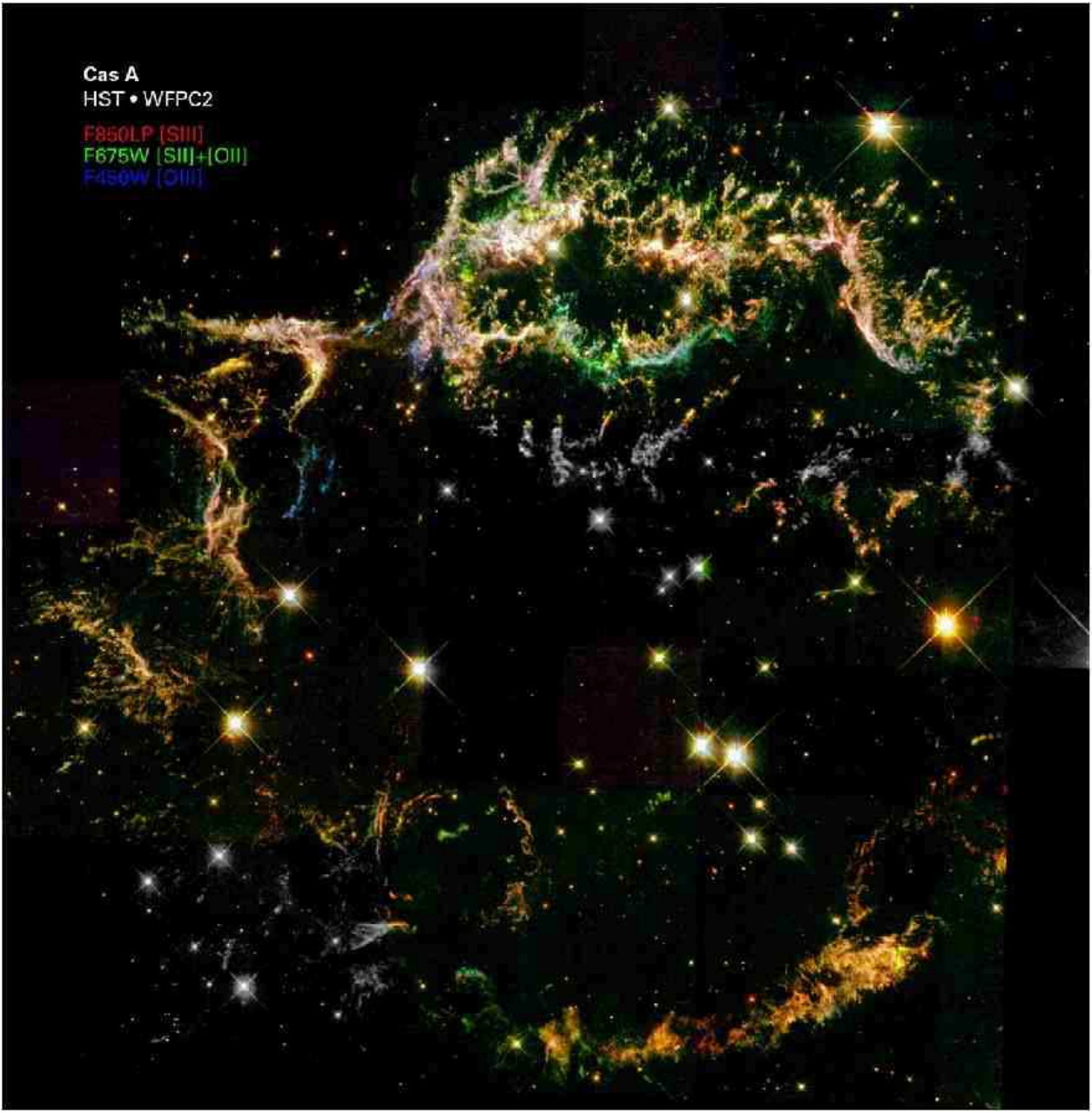}
\caption{WFPC-2 three color composite of the Cas A remnant.
The remnant's appearance in the F850LP (mainly [S~III] $\lambda\lambda$9069,9531 emission), F675W
(mainly [S~II] $\lambda\lambda$6716,6731, [O~II] $\lambda\lambda$7319,7330, and 
[O~I] $\lambda\lambda$6300,6364 emissions), and F450W (mainly [O~III] $\lambda\lambda$4959,5007 emission)  
filter images are shown in red, green, and blue, respectively. North is up and 
east is to
the left. The field-of-view shown is approximately 4 arc minutes on a side.
Areas with white only features are those imaged only in F675W (R band).
\label{fig1}}
\end{center}
\end{figure*}

\clearpage

\begin{figure*}
\begin{center}
\includegraphics[scale=0.79]{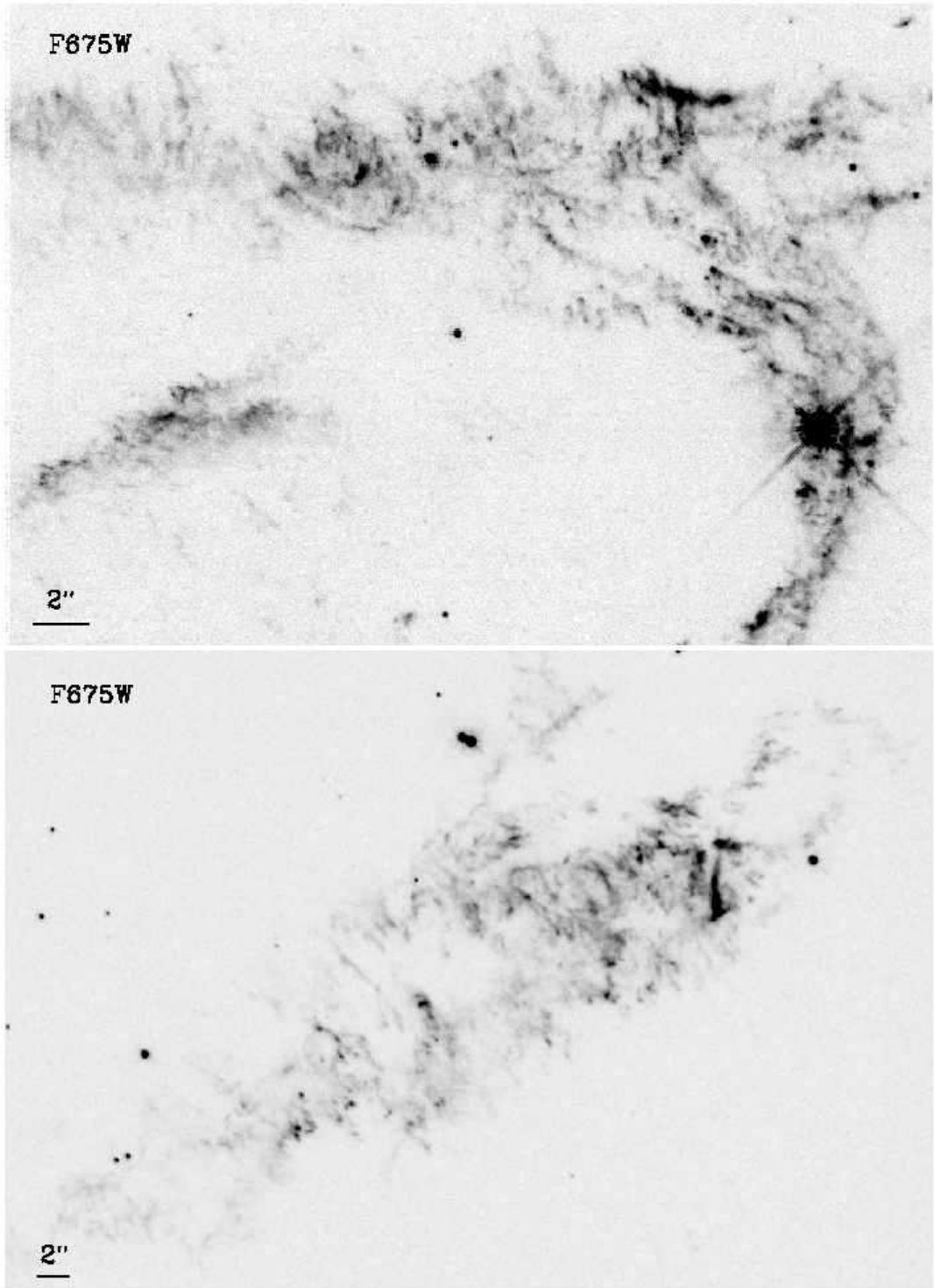}
\caption{Close-ups of the Baade-Minkowski Filament No. 1 (top; \citealt{BM54}) and the
remnant's bright southwestern rim region (bottom) as seen in the F675W filter.
\label{fig2}}
\end{center}
\end{figure*}

\clearpage

\begin{figure*}
\begin{center}
\includegraphics[scale=0.71,trim=0 0 10 0,clip=true]{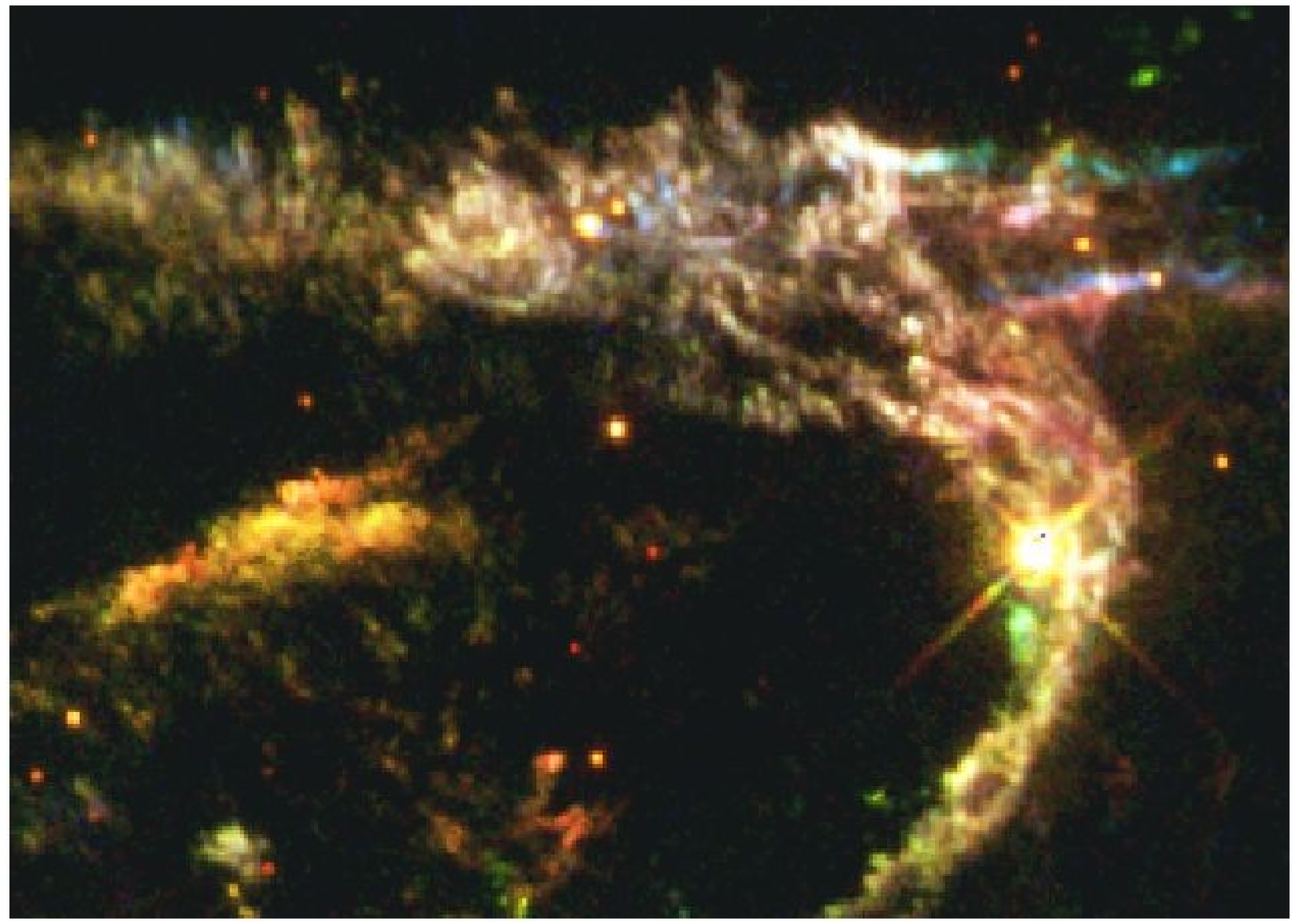}
\includegraphics[scale=0.7]{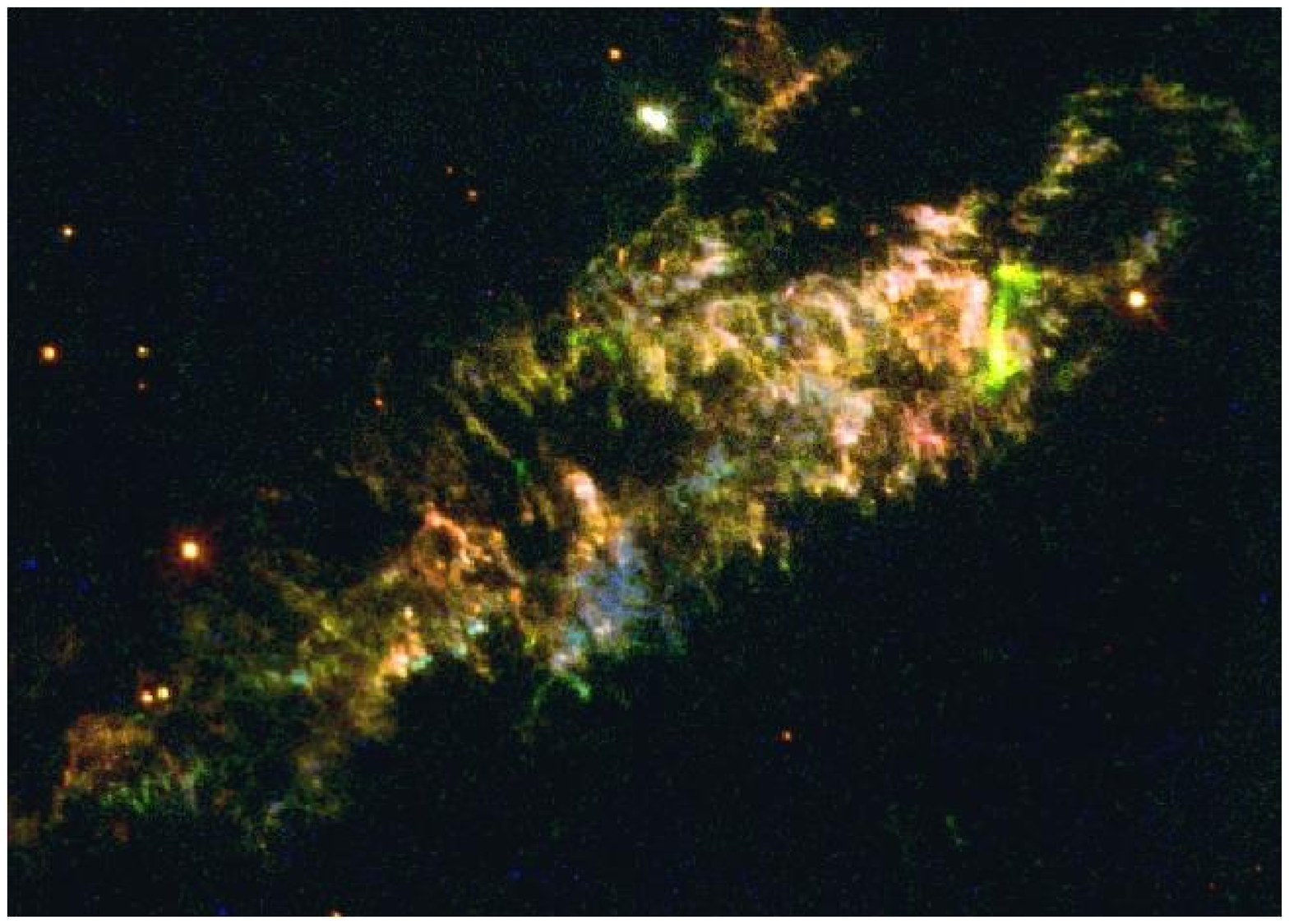}
\caption{False-color images of the Baade-Minkowski Filament No. 1 (top) and the
remnant's bright southwestern rim region (bottom). Color coding is the same
as used in Figure 1.
\label{fig3}}
\end{center}
\end{figure*}

\clearpage

\begin{figure*}
\begin{center}
\includegraphics[scale=0.85]{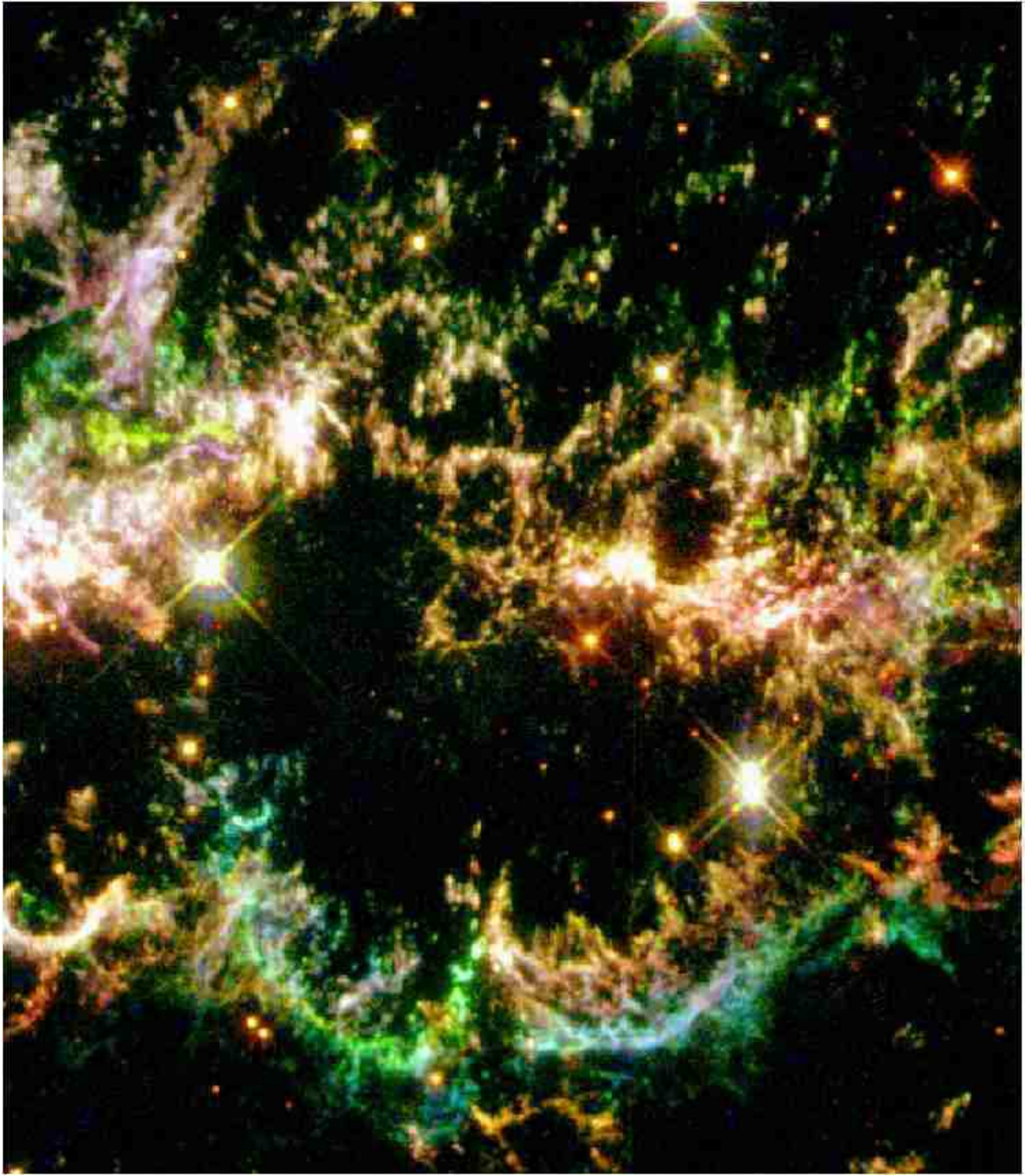}
\caption{False-color image of the remnant's northern limb.
Color coding is the same as used in Figure 1.
\label{fig4}}
\end{center}
\end{figure*}

\clearpage

\begin{figure*}
\begin{center}
\includegraphics[scale=0.85,clip=true,trim=0in 0in 0in 4in]{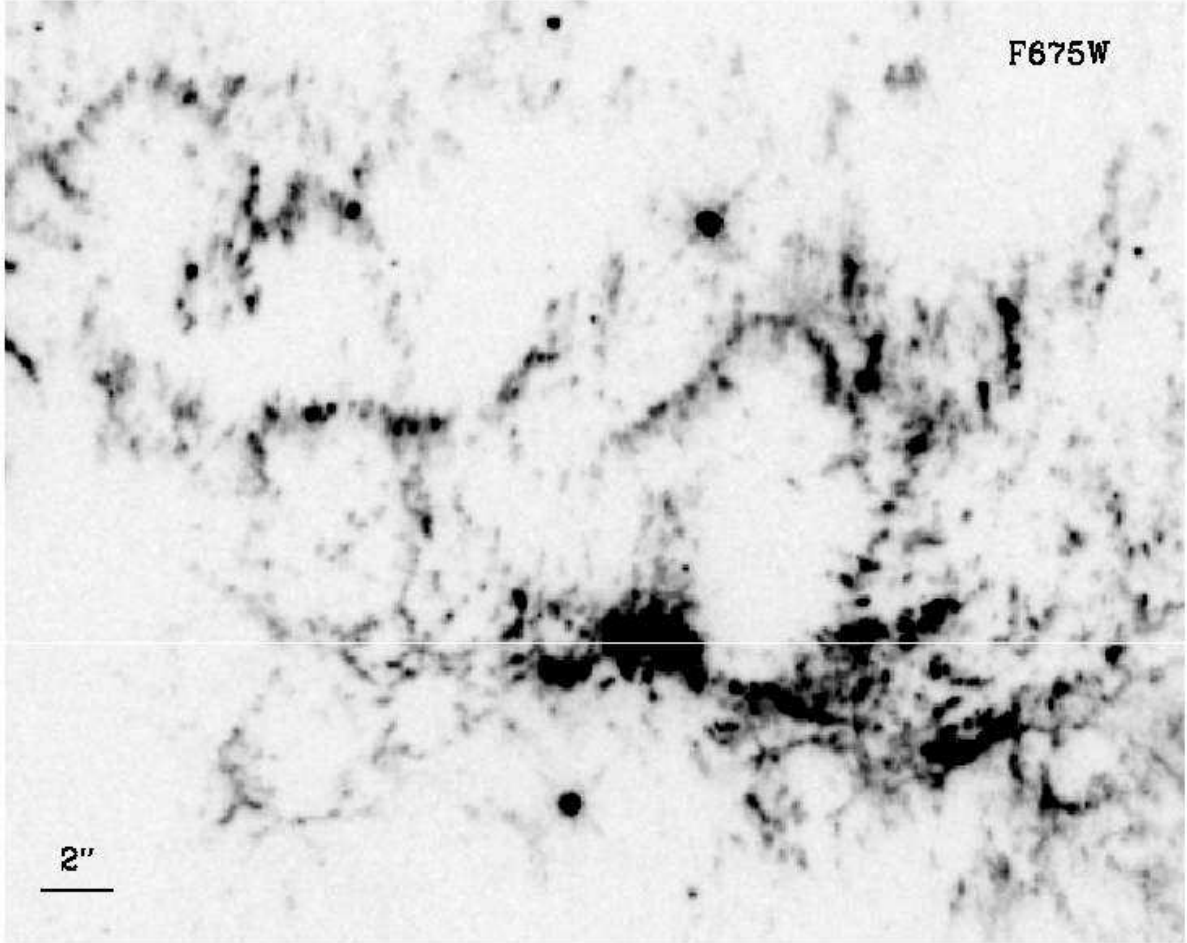}
\caption{F675W image enlargement of a portion of the north-central section of the bright shell showing
            numerous ejecta clumps.
\label{fig5}}
\end{center}
\end{figure*}

\clearpage

\begin{figure*}
\begin{center}
\includegraphics[scale=0.85]{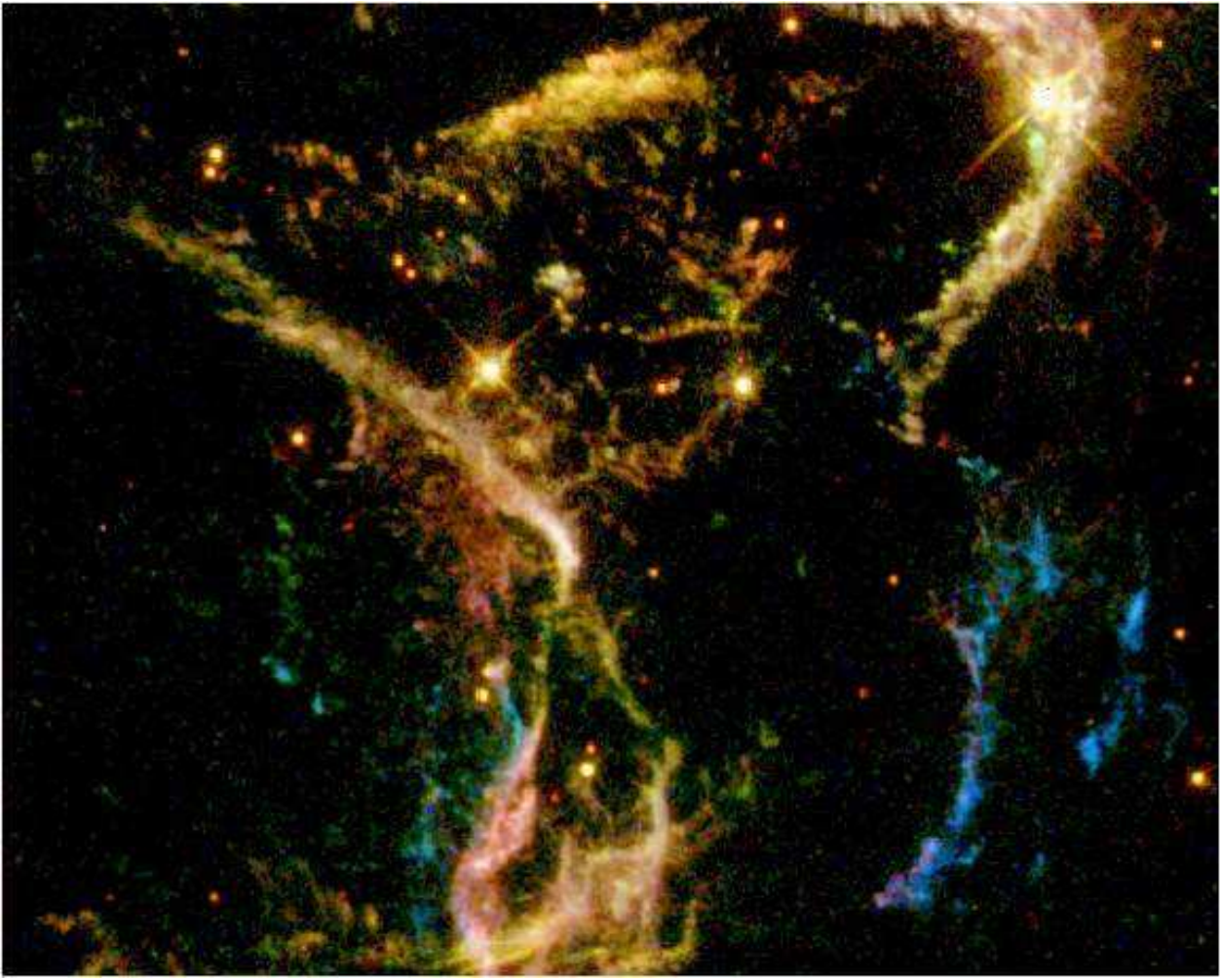}
\caption{False-color image of the remnant's northeastern limb near the base of the jet.
Color coding is the same as used in Figure 1.
\label{fig6}}
\end{center}
\end{figure*}

\clearpage

\begin{figure*}
\begin{center}
\includegraphics[scale=0.80]{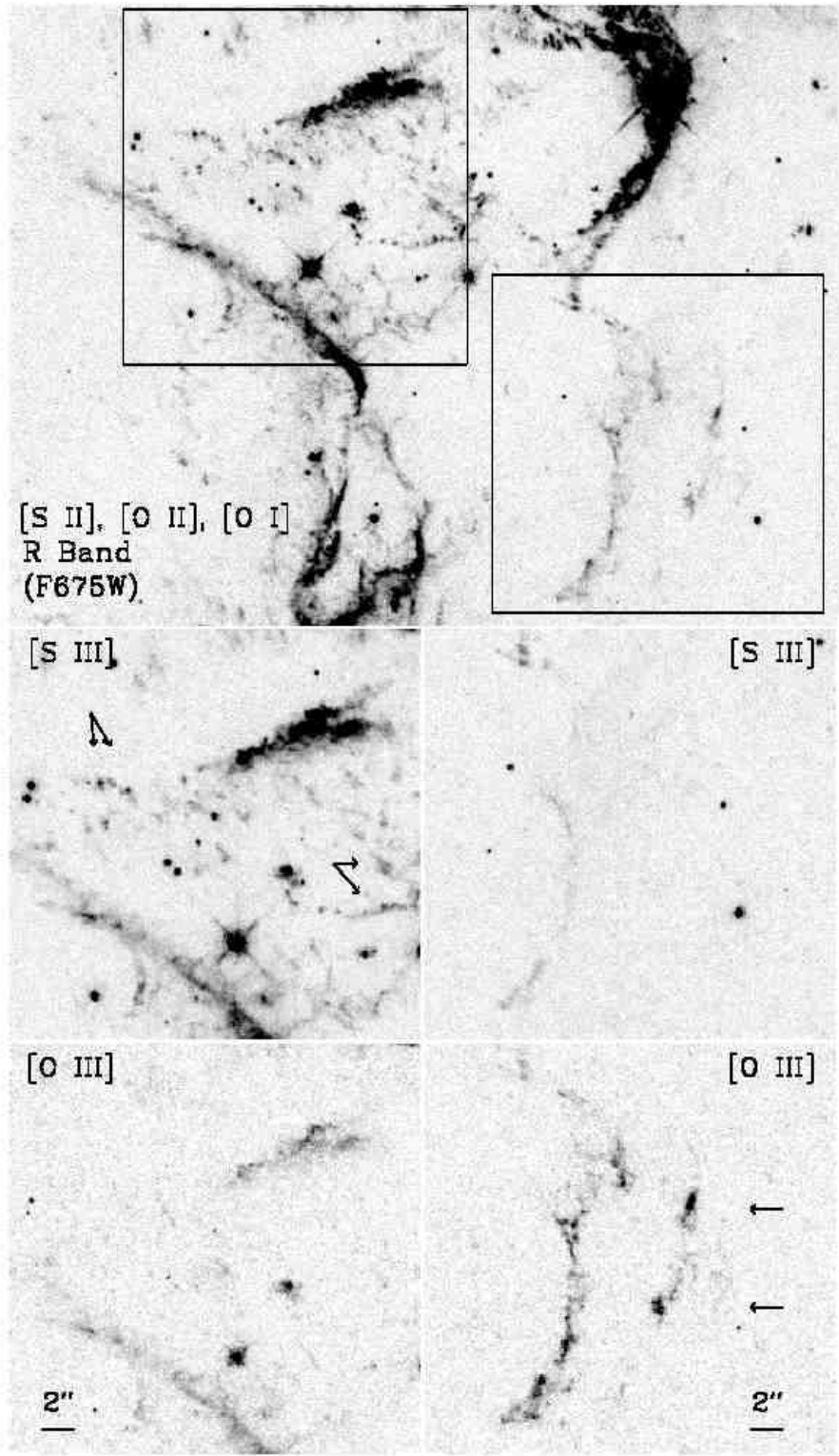}
\caption{Upper panel: F675W image of the NE region of the shell near the base of the jet.
     Marked regions are shown in enlargements below in [S~III] (F850LP) and [O~III] (F450W)
     sensitive images.
\label{fig7}}
\end{center}
\end{figure*}

\clearpage

\begin{figure*}
\begin{center}
\includegraphics[scale=0.80,trim=0in 0in 2in 0in,clip=true]{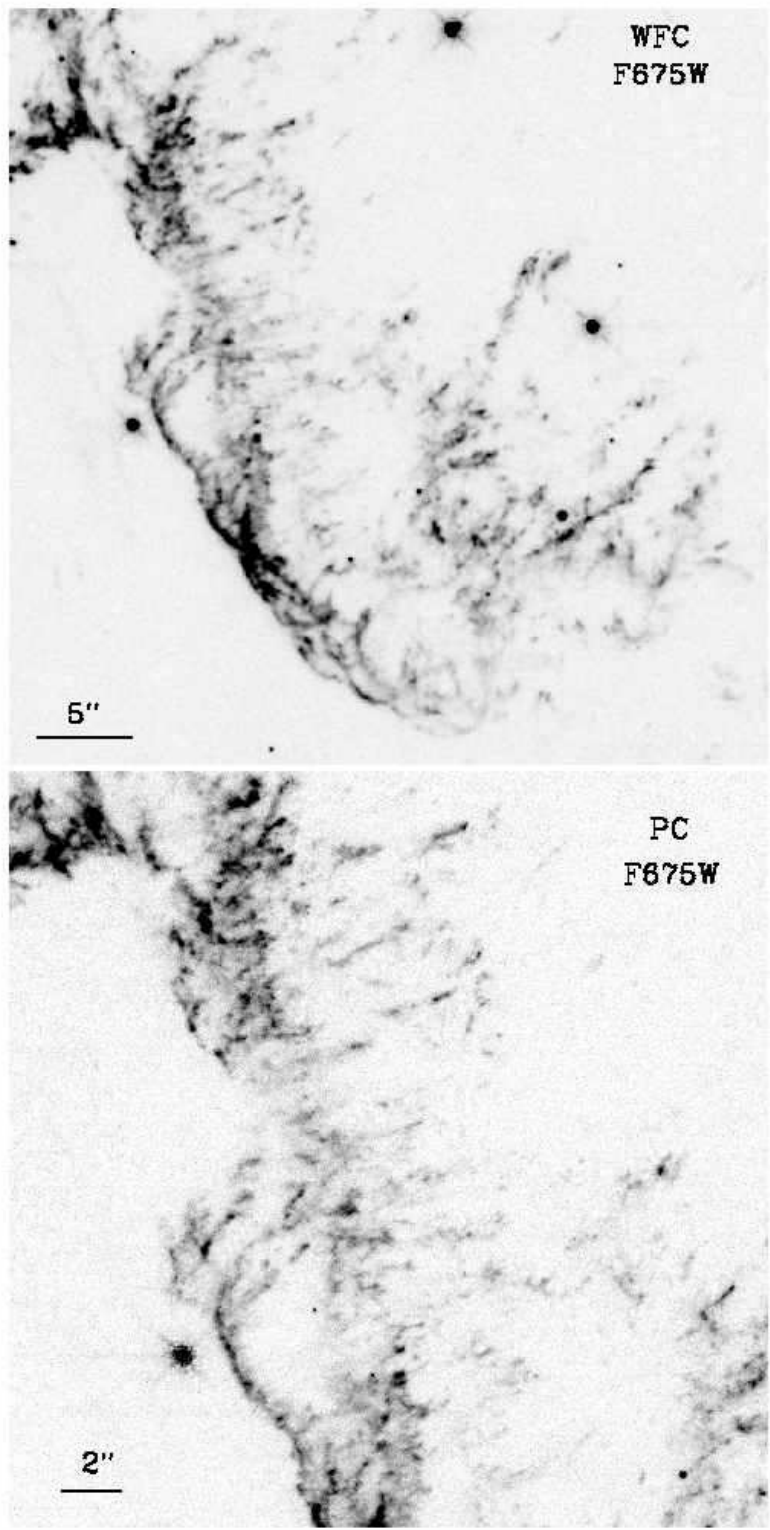}
\caption{A portion of a northwestern limb of the Cas~A shell showing
            the presence of the remnant's reverse shock front in this region.
            Lower panel shows a higher resolution PC image of a section of
            this area.
\label{fig8}}
\end{center}
\end{figure*}

\clearpage

\begin{figure*}
\begin{center}
\includegraphics[scale=0.8,clip=true,trim=0in 0in 0in 2in]{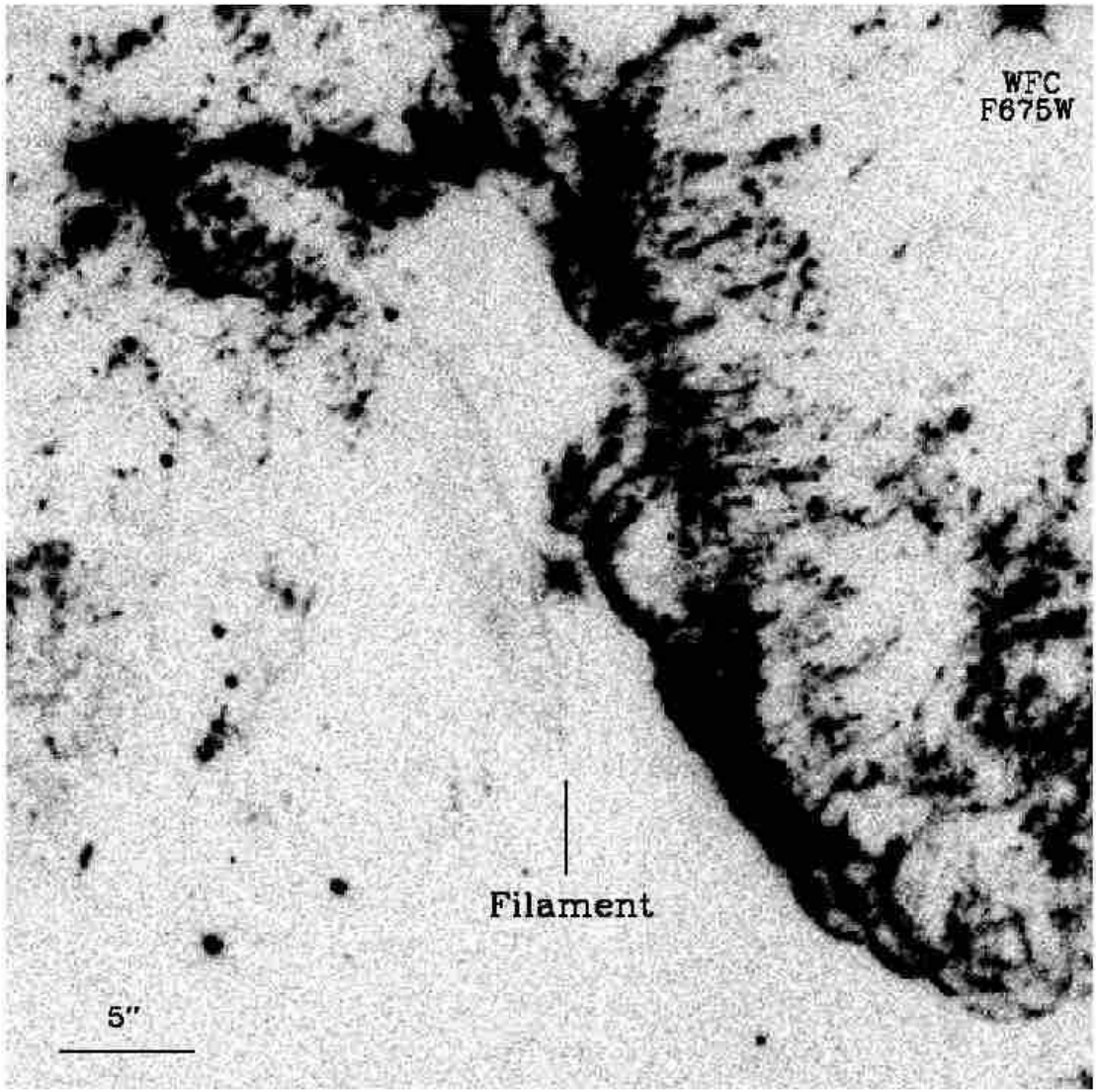}
\caption{F675W image of the northwest rim of the remnant showing a faint thin filament 
  east (left) of a line of suspected reverse shock front emission.  
\label{fig9}}
\end{center}
\end{figure*}

\clearpage

\begin{figure*}
\begin{center}
\includegraphics[scale=0.8,clip=true,trim=0in 5in 0in 0in]{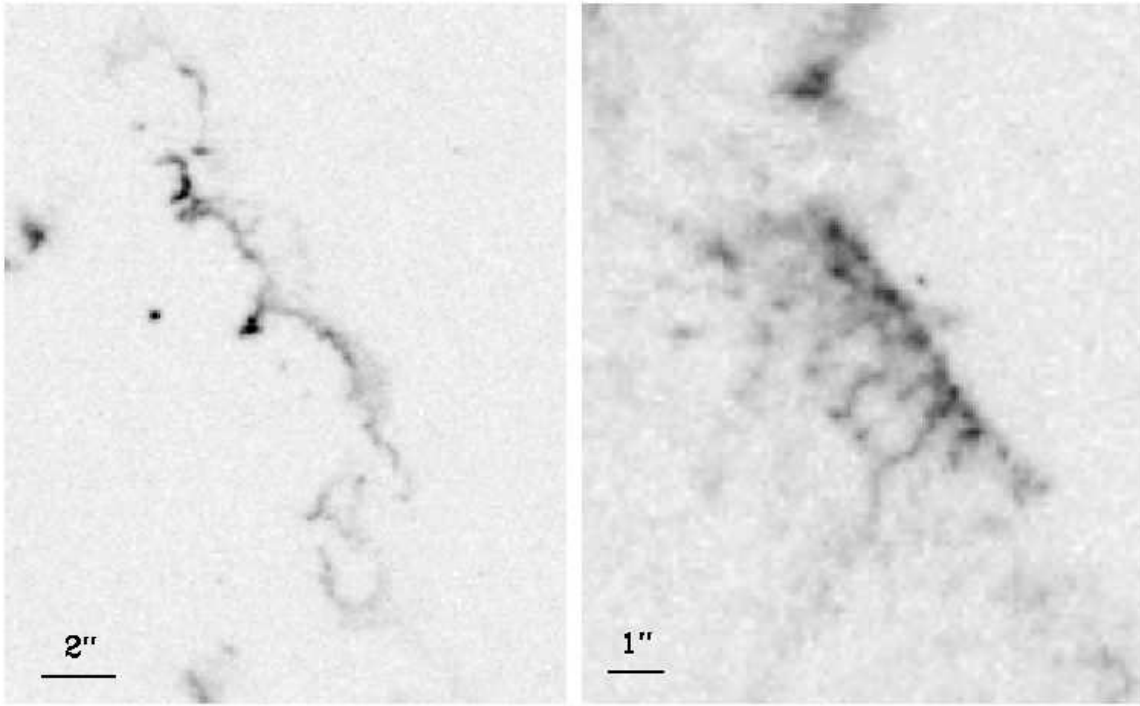}
\caption{F675W image enlargements of small knot structures in the south-central (left panel) 
            and eastern (right panel) regions of the Cas~A SNR.  
\label{fig10}}
\end{center}
\end{figure*}

\clearpage

\begin{figure*}
\begin{center}
\includegraphics[scale=0.78,clip=true,trim=0in 0.5in 0in 0in]{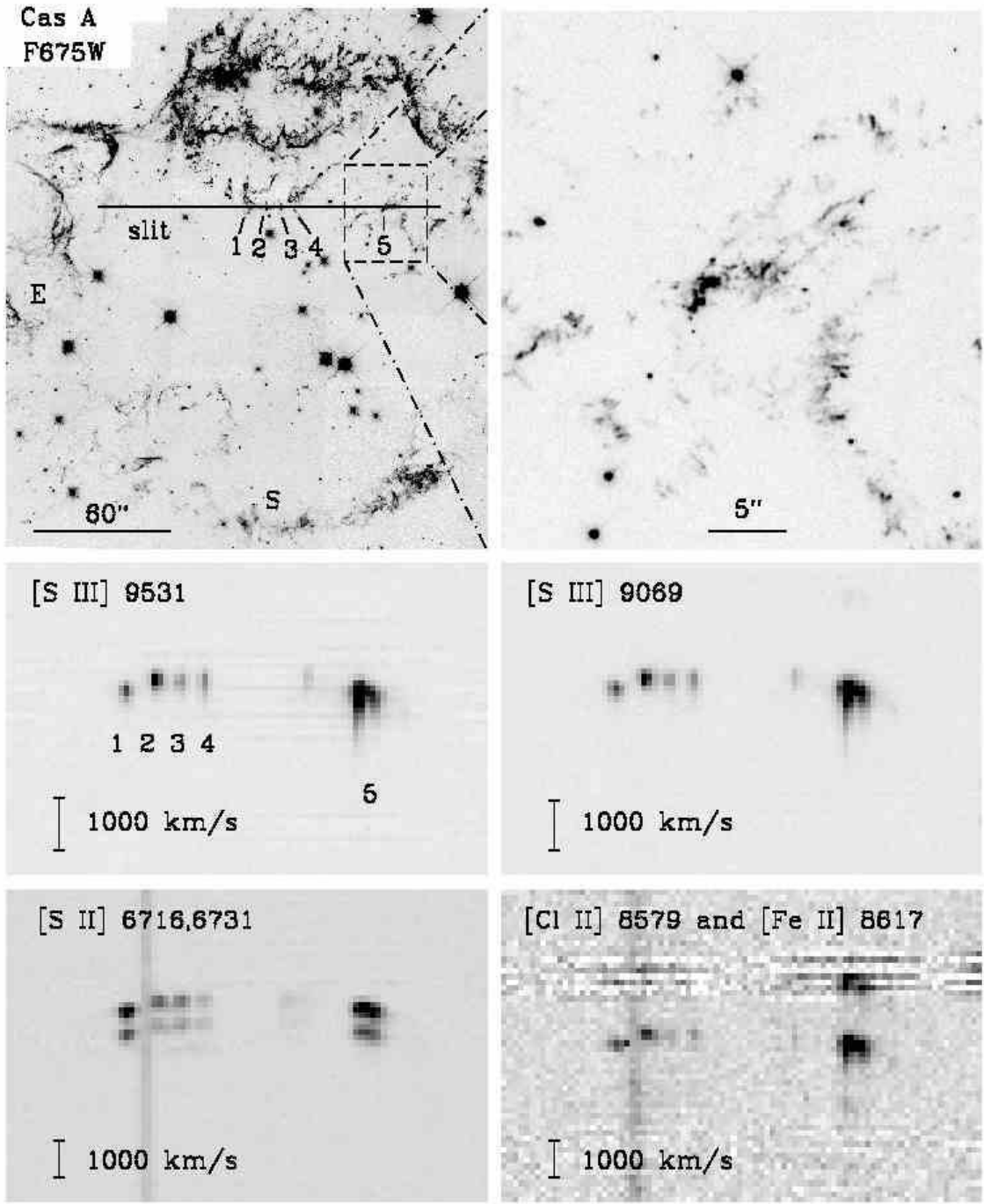}
\caption{Upper left: B\&W version of Fig. 1 showing slit location. Upper right: Close-up F675W image
    of nebulosity at and around Knot No.\ 5. Middle panels: Line profiles for [S~III] $\lambda\lambda$9069,9531.
    Lower panels: Emission line profiles for [S~II] $\lambda\lambda$6716,6731, [Cl~II] $\lambda$8579,
    and [Fe~II] $\lambda$8617. Wavelength increases towards the top of the panels with all
    five knots having radial velocities $\sim$+5000 km s$^{-1}$.
    Note the large ($\sim$ 1000 km s$^{-1}$) blueshifted emissions in the [S~III] and [Cl II] line
    profiles.  
\label{fig11}}
\end{center}
\end{figure*}

\clearpage

\begin{figure*}
\begin{center}
\includegraphics[scale=0.85,clip=true,trim=0in 1.5in 0in 0in]{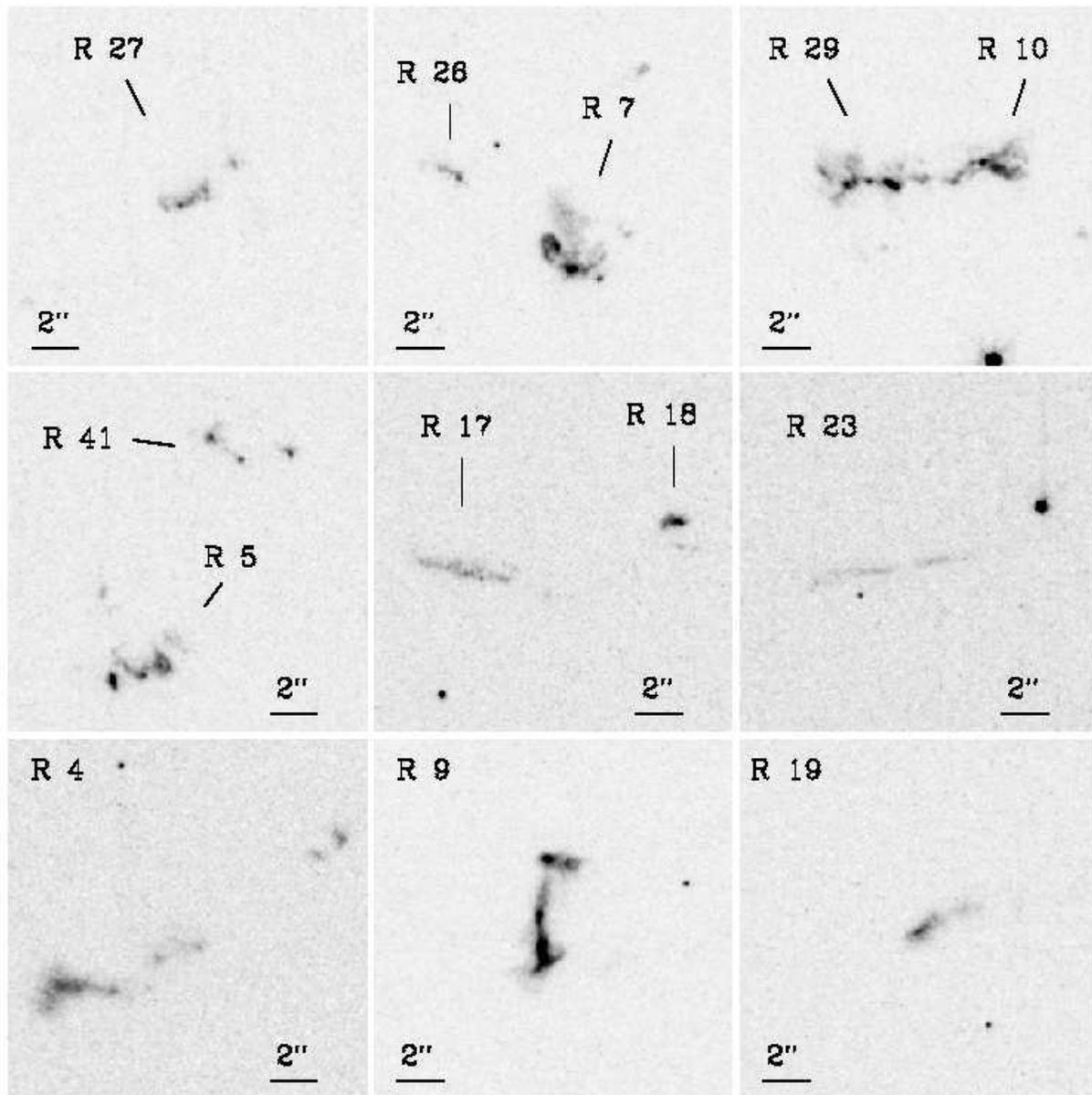}
\caption{[N~II] $\lambda$6583 line emission images of nine circumstellar knots (QSFs)
in the Cas~A SNR.
\label{fig12}}
\end{center}
\end{figure*}

\end{document}